\documentclass[arxiv,usenatbib]{iupartex}
\usepackage{newtxtext,newtxmath}
\usepackage[T1]{fontenc}

\title[Exploring the Eclipsing Binary System IY Aur]{Exploring the Eclipsing Binary System IY Aur: First Photometric Insights}
\author[Alan et al.]{
Neslihan Alan$^{1\cc}$,\orcid{0000-0001-9809-7493}
Mehmet Alpsoy$^{2}$\orcid{0009-0001-4009-7412},
and 
Y\"{u}cel Kılı\c{c}$^{3}$\orcid{0000-0001-8641-0796}
\affsep \\
$^1$Fatih Sultan Mehmet Vakif University, Department of History of Science, 34664, \.{I}stanbul, T\"{u}rkiye\\
$^2$Akdeniz University, Faculty of Science, Department of Space Sciences and Technologies, 07258, Antalya, T\"{u}rkiye\\
$^3$Instituto de Astrofísica de Andalucía (IAA-CSIC), Glorieta de la Astronomía s/n, 18008, Granada, Spain\\
}
\corres{N. Alan}{neslihan.alan@gmail.com}
\pubyear{2025}

\doiheader{XXXXXXX/PAR.20XX.00000}
\date{
	\pSubmit{00.00.0000} 
	\pRevReq{00.00.0000}
	\pLastRevRec{00.00.0000}
	\pAccept{00.00.0000}
	\pPubOnl{00.00.0000}
}
\volume{0}
\volnumber{1}

\begin{document}
\label{firstpage}
\pagerange{\pageref*{firstpage}--\pageref*{lastpage}}
\maketitle 

% Abstract of the paper
\begin{abstract}
Eclipsing binary systems play a vital role in astrophysics, as they provide a direct means of measuring fundamental stellar parameters. By combining high-precision space-based observations with ground-based multicolor photometric data, these parameters can be determined with greater accuracy. In this study, we present the first photometric analysis of the IY Aur eclipsing binary system, using a combination of the Transiting Exoplanet Survey Satellite ({\it TESS}) light curve and new {\it UBVRI} CCD observations obtained with the 60 cm robotic telescope (T60) at the T\"{U}B\.{I}TAK National Observatory. Through detailed photometric modeling, the masses and radii of the system’s primary and secondary components were determined as $M_{1}=6.51\pm 0.81\,M_\odot$, $M_{2}=5.39\pm0.87\,M_\odot$, and $R_{1}=4.15\pm 0.20\,R_\odot$, $R_{2}=6.88\pm 0.33\,R_\odot$, respectively. The logarithmic values of luminosity and surface gravity were calculated as $\log L$$_{1}$=3.14 $\pm$ 0.20 $L_\odot$ and $\log g_{1}$=4.01 $\pm$ 0.02 cgs for the primary component, and $\log L$$_{2}$=2.50 $\pm$ 0.22 $L_\odot$ and $\log g_{2}$=3.49 $\pm$ 0.03 cgs for the secondary component. Furthermore, the distance to IY Aur was estimated as $d=1690\pm237$ pc.
\end{abstract}

% Select between one and six entries from the list of approved keywords.
% Don't make up new ones.

%%%%%%%%%%%%%%%%%%%%%%%%%%%%%%%%%%%%%%%%%%%%%%%%%%

%%%%%%%%%%%%%%%%% BODY OF PAPER %%%%%%%%%%%%%%%%%%

\section{Introduction}

Eclipsing binary systems are essential tools in astronomical research, providing valuable insights into stellar structure, evolution, and galaxy dynamics. These systems allow the determination of fundamental stellar parameters directly through observations. The accuracy of these measurements is significantly enhanced by high-quality photometric data from space telescopes like {\it TESS} \citep{Ricker15}. Increased accuracy in determining these parameters contributes to testing theoretical models and improving stellar evolution theories by comparing evolutionary estimates with precise observational data.

Semi-detached binary systems, in which one component fills its Roche lobe and transfers mass to its companion, provide valuable opportunities to investigate mass transfer processes and the dynamics of interacting stellar systems. These interactions within semi-detached binaries can significantly affect the evolutionary paths of the component stars. The components of such systems evolve differently, and the mass transfer between them can alter their evolutionary paths in ways that detached systems cannot exhibit. By evaluating the fundamental parameters of these systems, it is possible to improve theoretical models of stellar evolution and to gain a deeper understanding of stellar structure, particularly in systems where one star is undergoing Roche lobe overflow.

IY Aur, a semi-detached binary system, was first reported as a variable star by \citet{Weber63}. Some general information on the IY Aur system from the General Catalogue of Variable Stars \citep[GCVS,][]{Samus17} is given in Table~\ref{general_parameters}. Recently, \citet{Khalatyan24} utilized the {\it Gaia} DR3 dataset, particularly the low-resolution XP spectra, to derive the atmospheric parameters and masses of 217 million stars, including IY Aur. For the IY Aur system, the mean values of the surface gravity ($\log g$), effective temperature ($\log T_{\rm eff}$), mass, and metallicity were determined to be 3.663$\pm$0.368 cgs, 4.238$\pm$0.040 K, 6.512$\pm$0.0813 $M_\odot$, and -0.872$\pm$0.501 dex, respectively. Despite these broad surveys, which provide stellar atmosphere model parameters and mass estimates for the system, no detailed spectral or photometric analyses of IY Aur have been conducted up to now. In this context, there is a notable gap in the literature concerning the IY Aur system. This absence underscores the need for targeted observations and analyses to determine the nature of IY Aur and its components. The IY Aur system, exhibiting periodic eclipses, offers an exciting opportunity for detailed photometric and spectroscopic investigations to refine our understanding of its stellar properties and evolutionary stage.

% TABLE 1 
\begin{table} 
\setlength{\tabcolsep}{5mm} % adjust space between the table columns
\renewcommand{\arraystretch}{1.0} % adjust space between the table rows
\centering
\caption{GCVS information of IY Aur.}
\begin{tabular}{lrrr}
\hline
 Parameter			 			               & Value		\\	
\hline
RA  (J2000) & $05^{\rm h} 48^{\rm m} 27^{\rm s}.20$\\
DEC (J2000) & $+43^{\circ} 04^{'} 57^{''}.50$\\
$V_{\rm max}$ ${\rm (mag)}$ & 9.270 \\
$P$ (day) & 2.79338 \\
Type of Variability & EB\\ 
Spectral Type & B5:p\\ 
 \hline
\end{tabular}
\label{general_parameters}
\end{table}

In this study, we aim to determine the fundamental stellar parameters of the components of the IY Aur system. To achieve this, the first detailed light curve analysis of the system was carried out based on photometric data obtained from {\it TESS} observations and ground-based T60 telescope measurements. As a result of this analysis, the fundamental parameters of both components were determined with reasonable accuracy. The structure of this paper is organized as follows: Section 2 describes the observational data used in this research and the methodology adopted for data reduction. In Section 3, the procedure for the simultaneous analysis of the {\it TESS} data combined with ground-based photometric observations is presented in detail. Section 4 reports the fundamental stellar parameters of IY Aur derived from the light curve analysis. Finally, Section 5 discusses the obtained results and presents an overall assessment of the study.

\section{PHOTOMETRIC DATA}

Over the course of 101 nights, from November 1, 2018, to January 5, 2022, new multicolour CCD observations of IY Aur were conducted using the 60 cm robotic telescope (T60) at the T\"UB\.ITAK National Observatory (TUG). The T60 telescope is operated via the open-source OCAAS software, officially referred to as TALON \citep{Parmaksizoglu14}. Initially, observations were carried out using the FLI ProLine 3041-UV CCD camera until July 23, 2019, after which the Andor iKon-L 936 BEX2-DD camera was employed. The FLI ProLine 3041-UV CCD provides an image scale of 0.51 arcseconds per pixel, resulting in a field of view (FOV) of 17.4 arcminutes, whereas the Andor iKon-L 936 BEX2-DD camera offers a slightly finer image scale of 0.456 arcseconds per pixel, corresponding to a FOV of 15.6 arcminutes.

\begin{figure}
	\begin{center}
		\label{fig1}
		\includegraphics[width=0.92\columnwidth]{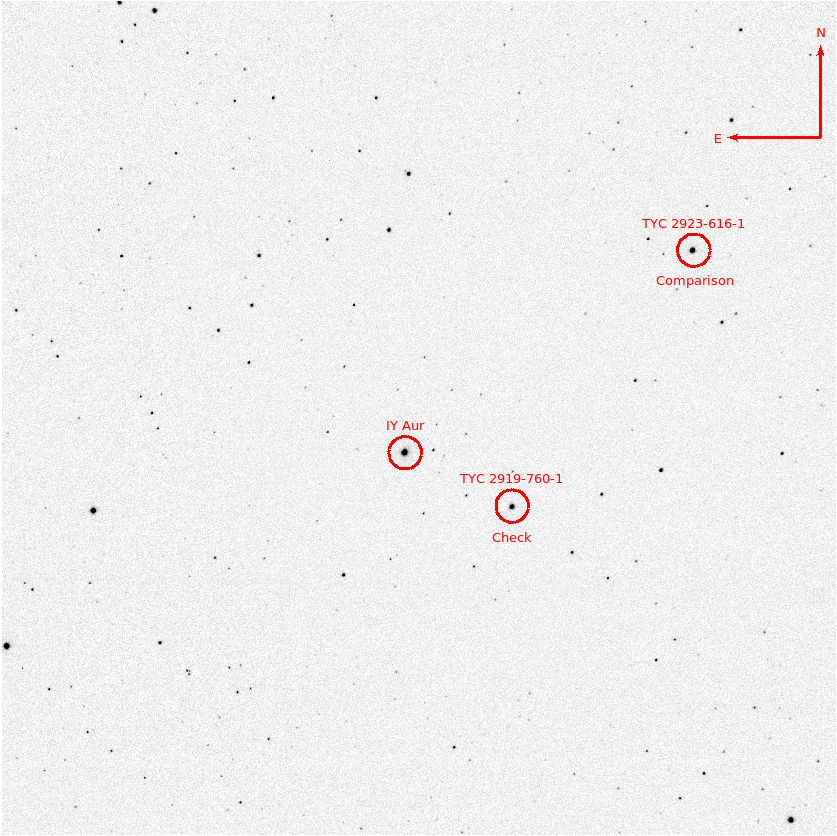}
		\caption{An image taken with the $V$ filter from the TUG-T60 telescope, with labels indicating the comparison and check stars, as well as IY Aur.}
        \label{fig:star_field}
	\end{center}
\end{figure}

Observations of IY Aur were conducted using Bessell {\it UBVRI} filters \citep{Bessell90}, with exposure times of 60 s for {\it U}, and 5s each for the $B$, $V$, $R$, and $I$ filters. To account for pixel-to-pixel variations in the CCD sensor, calibration frames—including sky flats and bias frames—were regularly acquired. For differential photometry, TYC 2923-616-1 was chosen as the comparison star, while TYC 2919-760-1 served as the check star, ensuring the accuracy and consistency of the measurements. An image captured by the T60 telescope using the {\it V} filter is shown in Figure~\ref{fig:star_field}.

\begin{figure}[!ht]
\begin{center}
\includegraphics*[scale=0.65,angle=000]{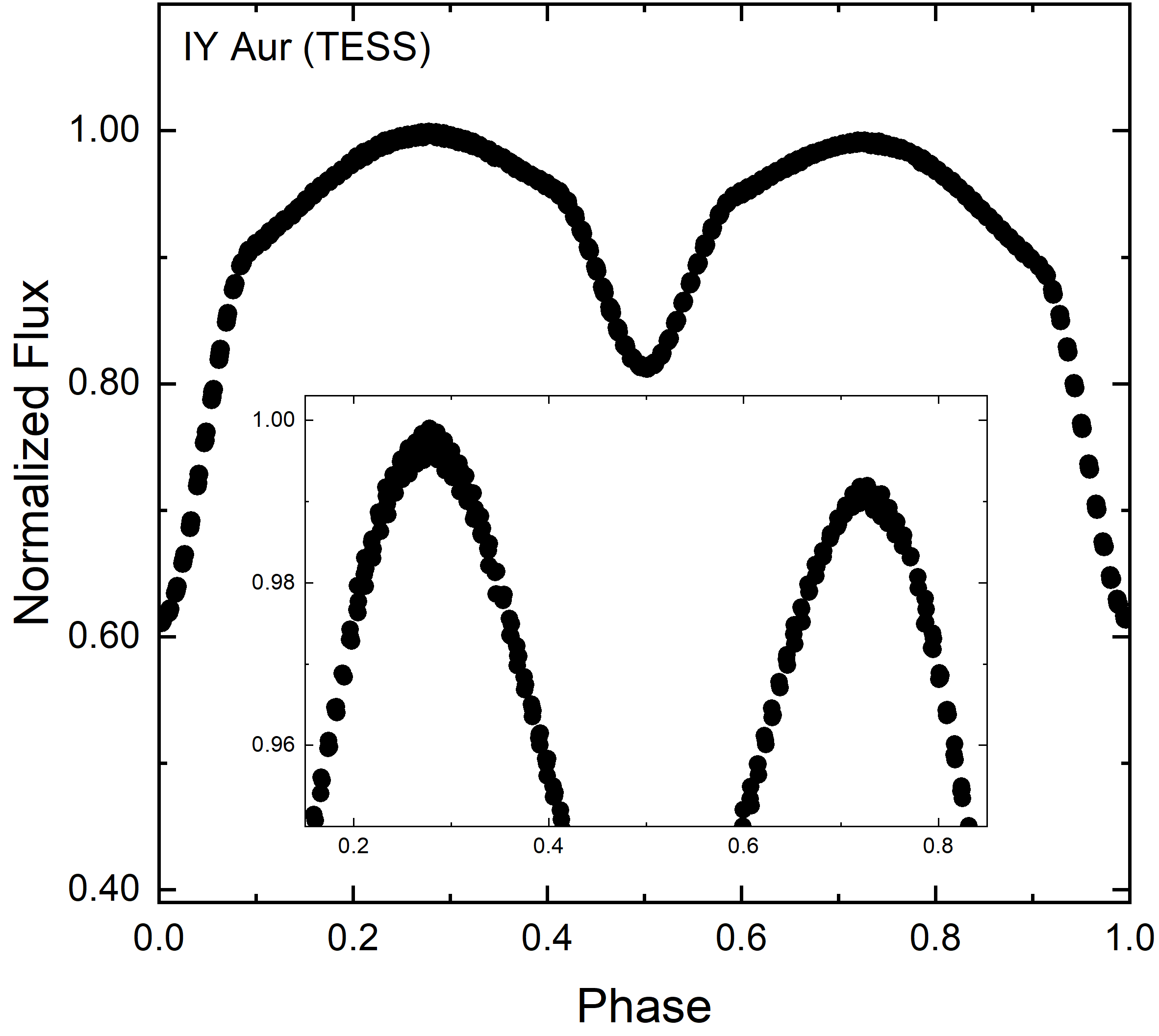}
\caption{A phased {\it TESS} light curve. The zoomed-in panel displays noticeable differences in flux levels between phases 0.25 and 0.75.} \label{fig:Tess_LC_fit}
\end{center}
\end{figure}

In addition to ground-based observations, the light curve analysis incorporates data from the Transiting Exoplanet Survey Satellite ({\it TESS}), which systematically scans large portions of the sky in 27.4-day observational sectors. {\it TESS} operates within a wavelength range of 600–1000 nm, delivering broadband photometric data \citep{Ricker15}. Observations of IY Aur were carried out in Sector 19 from November 28 to December 23, 2019, with an exposure time of 1800 seconds. The system’s {\it TESS} data were retrieved from the Mikulski Archive for Space Telescopes (MAST)\footnote{https://archive.stsci.edu/}. For analysis, Pre-search Data Conditioning (PDC) Simple Aperture Photometry (SAP) light curves were utilized, as outlined by \citet{Ricker15}. The average photometric uncertainty in the dataset is approximately 0.1\%. Light curves derived from the high-precision {\it TESS} observation data exhibit out-of-eclipse light variations. Figure~\ref{fig:Tess_LC_fit} shows a sample phased {\it TESS} light curve, with the inset panel providing a detailed view of the flux variations at phases 0.25 and 0.75.

%Figure 1

%Figure 2

The data reduction process for the T60 observations followed a series of essential steps. Initially, the science frames underwent bias and dark frame subtraction, followed by flat-field correction. The processed CCD images obtained from these steps were then used to determine the differential magnitudes of the target stars. This analysis was carried out using MYRaf, an IRAF-based aperture photometry GUI tool developed by \citet{Kilic16}. Throughout the observation period, no significant variations were detected in the light curves of the comparison and check stars. The external uncertainties in the magnitude differences between these stars were calculated as 39 mmag in the $U$ filter, 38 mmag in the $B$ filter, 28 mmag in the $V$ filter, 30 mmag in the $R$ filter, and 30 mmag in the $I$ filter. These values were derived from the standard deviation of the differential magnitudes recorded over the same night. Instead of transforming the observational data into the standard Bessell {\it UBVRI} system, the differential magnitudes were directly utilized for light curve analysis.

\section{SIMULTANEOUS ANALYSIS OF MULTICOLOUR LIGHT CURVES}

\begin{figure}[!ht]
\begin{center}
\includegraphics*[scale=0.65,angle=000]{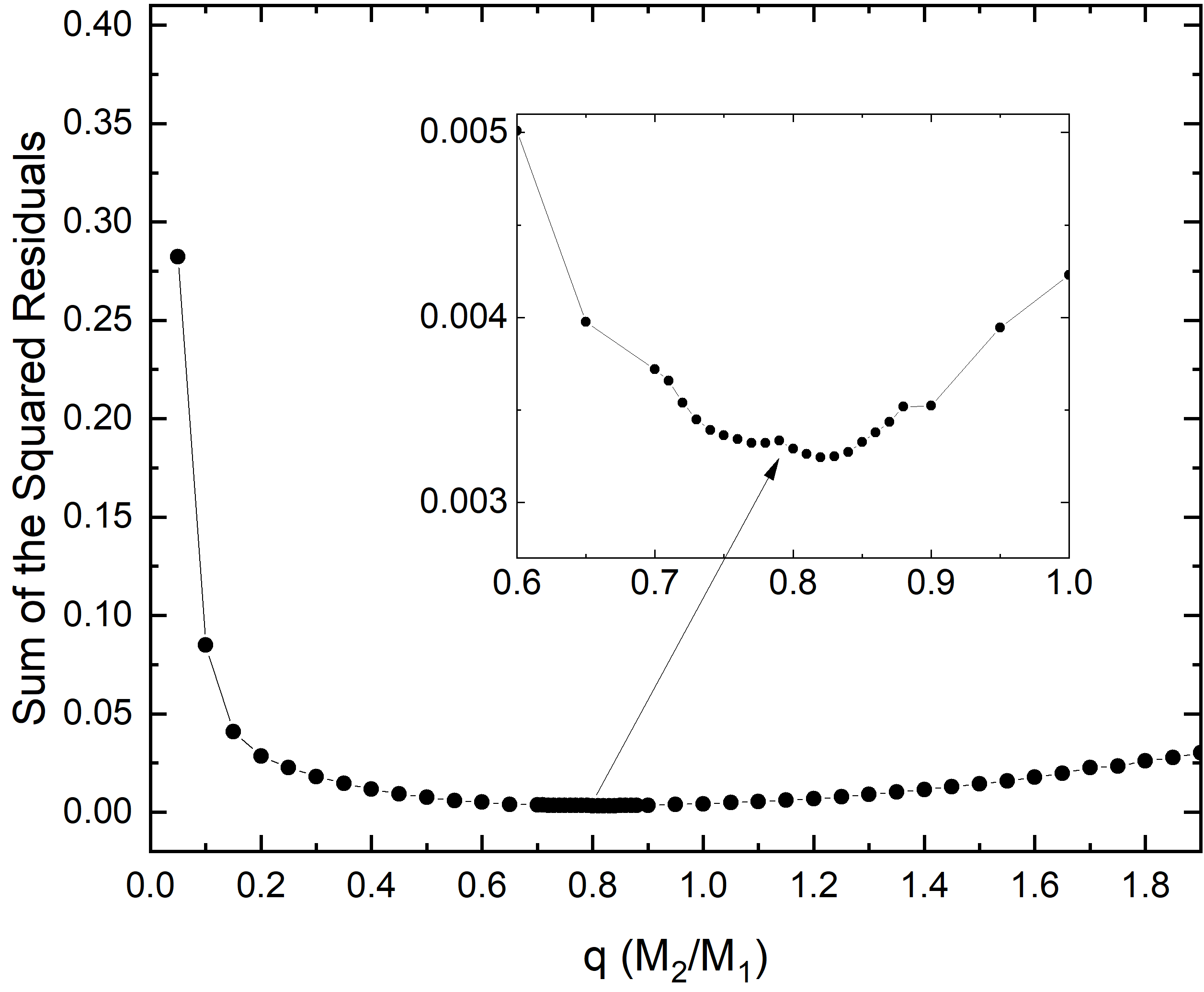}
\caption{Illustration of the $q$-search analysis results for IY Aur.} \label{fig:Q_search}
\end{center}
\end{figure}

A comprehensive photometric analysis of the IY Aur system's light curve was conducted using both normalized ground-based $UBVRI$ observations and {\it TESS} data simultaneously. The analysis employed the Wilson-Devinney (W--D) method \citep{WilsonDevinney71}, and the uncertainties in the resulting parameters were evaluated using Monte Carlo simulations \citep{Zola04, Zola10}. Since spectroscopic radial velocity measurements were not available for IY Aur, the mass ratio ($q$) was determined using the $q$-search technique \citep{Terrell05}, which relies on light curve modeling. This method, known as the photometric mass ratio test, minimizes the chi-square ($\chi^2$) statistic through iterative calculations to find the most likely $q$ value. During the $q$-search, all parameters were kept free, while $q$ was incrementally fixed at 0.01 intervals between 0 and 3. The analysis revealed that the minimum $\chi^2$ value occurred at approximately $q=0.82$. The outcomes of the $q$-search performed on IY Aur are presented in Figure~\ref{fig:Q_search}. IY Aur has been classified in the General Catalogue of Variable Stars (GCVS) as an EB-type eclipsing binary system \citep{Samus17}. The light curve solution was initially attempted using the MOD 2 configuration, in which both components are within their Roche lobes; however, this approach did not produce physically meaningful results. 
Subsequently, the solution was performed via MOD 4, in which the primary component fills its Roche lobe, yet an acceptable fit could not be achieved. Therefore, the modeling was repeated using MOD 5, where the secondary component fills its Roche lobe, and theoretical models were generated that successfully represent the observational data.

\begin{figure}[!ht]
\begin{center}
\includegraphics*[scale=0.65,angle=000]{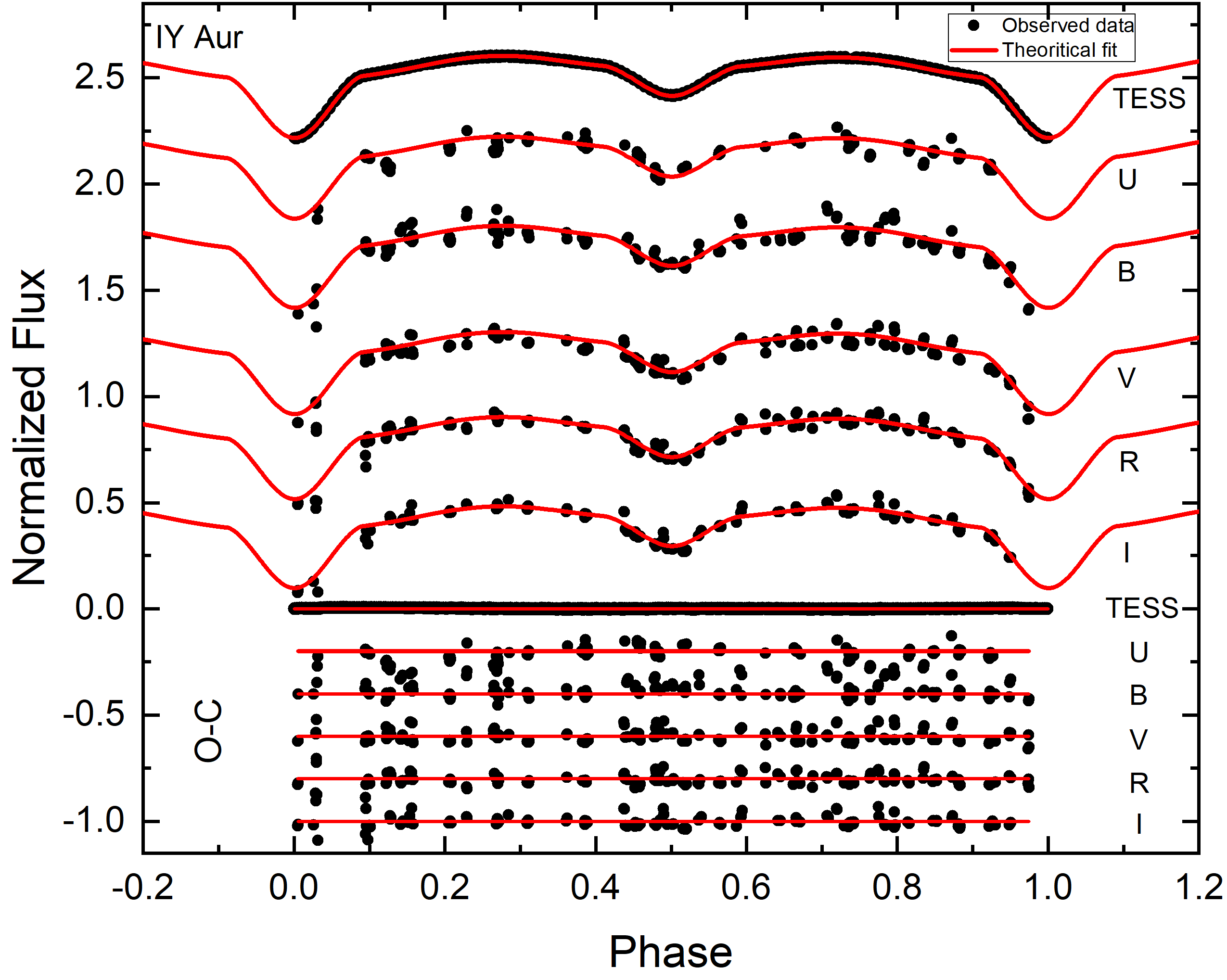}
\caption{A comparison between the theoretical light curves (red line) and the observed data (black dots) for IY Aur. The residuals are shown in the bottom panel.} \label{fig:LC_fit}
\end{center}
\end{figure}

\begin{figure}[!ht]
\begin{center}
\includegraphics*[scale=0.55,angle=000]{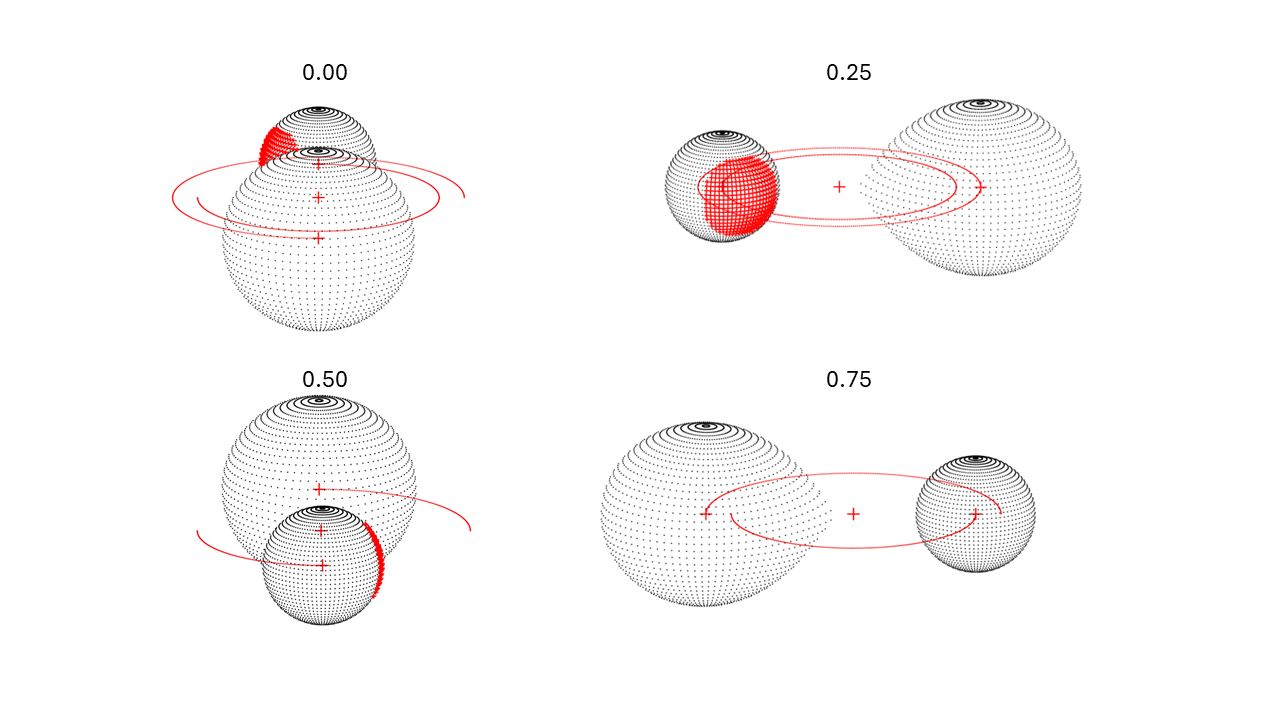}
\caption{The Roche geometry of the IY Aur system in four different phases was constructed based on the best-fit light curve model parameters.} \label{fig:Roche_geometry}
\end{center}
\end{figure}

The W--D code analysis was carried out by combining fixed parameters, based on literature and theoretical models, with adjustable parameters that were refined through multiple iterations. The fixed and adjustable parameters are detailed below. The effective temperature of the primary component was fixed at 17290 K, as provided by \citet{Khalatyan24} using {\it Gaia} DR3 data. The root-mean-square limb darkening law was applied, using limb darkening coefficients from \citet{Vanhamme93} based on the temperatures of the IY Aur components and filter wavelengths. For both components with radiative atmosphere ($T_{\rm eff}>7200$ K), the bolometric gravity-darkening exponents were fixed at 1 \citep{Lucy67}, while the bolometric albedos were also set at 1 according to \citet{Rucinski69}. It was assumed that both components are in synchronous rotation, with rotational parameters set to $F_{\rm 1}=F_{\rm 2}=1$. Orbital eccentricity ($e$) was assumed to be zero. In the analysis, the orbital inclination ($i$), mass ratio ($q$), the effective temperature of the secondary component ($T_{\rm 2,eff}$), the dimensionless surface potential of the primary component ($\Omega_{1}$), phase shift, and the primary component's fractional luminosity ($L_{1}$) were treated as free parameters. To investigate the potential presence of a third body, the third light contribution ($l_{3}$) was included as a free parameter, but no evidence of third light was found for IY Aur. The light curve solution was initially performed without considering the presence of any spots. However, this model was unable to adequately represent the significant out-of-eclipse light variations. Therefore, models containing spots have been applied in order to model these out-of-eclipse light variations. The best fit was achieved by adopting an approach that assumes the presence of a hot spot on the primary component. A statistically notable improvement is indicated by the lower chi-square value of the spotted model ($\chi^{2}=0.003$) compared to the spotless model ($\chi^{2}=0.008$). The spotted model effectively explains the out-of-eclipse light variations and shows strong agreement with the observations. According to the spotted model, the hot spot on the surface of the primary component was located at a co-latitude of $90^\circ$ and a longitude of $297^\circ$, with an angular radius of $49^\circ$. The relative temperature parameter of the spot was calculated as 1.03. The parameters derived from both the spotted and spotless models are summarized in Table~\ref{tab:lcparameters}. Figure~\ref{fig:LC_fit} presents a comparison between the theoretical and observed light curves based on the spotted model, and Figure~\ref{fig:Roche_geometry} shows the corresponding Roche geometry of IY Aur.  

% TABLE 2 

\begin{table}
 \setlength{\tabcolsep}{5mm} % adjust space between the table columns
\renewcommand{\arraystretch}{1.0} % adjust space between the table rows
\begin{center}
\centering
\caption{The results of the light curve analysis for IY Aur. The subscripts 1, 2, and 3 correspond to the primary, secondary, and third components, respectively. Parameters marked with $^{\rm a}$ indicate fixed values.}
\begin{tabular}{lrrr}
\hline
 Parameter			 			                           & \multicolumn{2}{c}{Value} \\		
                                                             & Spotless & Spotted
 \\
\hline
$T_{0}$$^{\rm a}$  {\rm (BJD+2400000)}          & \multicolumn{2}{c}{52502.0956} \\
$P_{\rm orb}$$^{\rm a}$  (days)                 & \multicolumn{2}{c}{2.7933774} \\
$i$ ($^{\rm o}$)	       	                    & 74.076 $\pm$ 0.016 &  73.835 $\pm$ 0.018  \\	
$T$$_{\rm 1,eff}$$^{\rm a}$ (K)                 & \multicolumn{2}{c}{17290} 	\\	
$T$$_{\rm 2,eff}$ (K)    	          		    & 9372 $\pm$ 1621 & 9293 $\pm$ 1615		\\
$e$$^{\rm a}$ 	         	               	    & \multicolumn{2}{c}{0.000} 		  \\
$\Omega$$_{1}$		           	                & 5.463 $\pm$ 0.019 & 5.443 $\pm$ 0.014 	  \\ 
$\Omega$$_{2}$		            	            & 3.441 $\pm$ 0.046 & 3.465 $\pm$ 0.041 	  \\
Phase shift             	  	                & 0.0004 $\pm$ 0.0001 & 0.0010 $\pm$ 0.0001	  \\
$q$                     	  	                & 0.814 $\pm$ 0.034  & 0.828 $\pm$ 0.030	  \\
$r$$_{\rm 1}^*$ (mean)                          & 0.216 $\pm$ 0.001  & 0.218 $\pm$ 0.001  \\
$r$$_{\rm 2}^*$ (mean)                          & 0.360 $\pm$ 0.001  & 0.361 $\pm$ 0.001  \\
$L$$_{\rm 1}$/($L$$_{1}$+$L$$_{2}$) ($TESS$) 	& 0.452 $\pm$ 0.001 & 0.458 $\pm$ 0.001   \\
$L$$_{\rm 1}$/($L$$_{1}$+$L$$_{2}$) ($U$)   	& 0.767 $\pm$ 0.008 & 0.774 $\pm$ 0.008   \\
$L$$_{\rm 1}$/($L$$_{1}$+$L$$_{2}$) ($B$)   	& 0.574 $\pm$ 0.006 & 0.581 $\pm$ 0.006   \\
$L$$_{\rm 1}$/($L$$_{1}$+$L$$_{2}$) ($V$)   	& 0.523 $\pm$ 0.006 & 0.530 $\pm$ 0.006   \\
$L$$_{\rm 1}$/($L$$_{1}$+$L$$_{2}$) ($R$)  	    & 0.492 $\pm$ 0.005 & 0.499 $\pm$ 0.005   \\
$L$$_{\rm 1}$/($L$$_{1}$+$L$$_{2}$) ($I$)   	& 0.452 $\pm$ 0.005 & 0.459 $\pm$ 0.005   \\
$L$$_{\rm 2}$/($L$$_{1}$+$L$$_{2}$) ($TESS$)    & 0.548 $\pm$ 0.001 & 0.542 $\pm$ 0.001   \\
$L$$_{\rm 2}$/($L$$_{1}$+$L$$_{2}$) ($U$)   	& 0.233 $\pm$ 0.003 & 0.226 $\pm$ 0.003   \\
$L$$_{\rm 2}$/($L$$_{1}$+$L$$_{2}$) ($B$)   	& 0.426 $\pm$ 0.002 & 0.419 $\pm$ 0.002   \\
$L$$_{\rm 2}$/($L$$_{1}$+$L$$_{2}$) ($V$)  	    & 0.477 $\pm$ 0.002 & 0.470 $\pm$ 0.002   \\
$L$$_{\rm 2}$/($L$$_{1}$+$L$$_{2}$) ($R$)   	& 0.508 $\pm$ 0.002 & 0.501 $\pm$ 0.002   \\
$L$$_{\rm 2}$/($L$$_{1}$+$L$$_{2}$) ($I$)   	& 0.548 $\pm$ 0.002 & 0.541 $\pm$ 0.002   \\
$l$$_{\rm 3}$ ($TESS, U, B, V, R, I$)           & \multicolumn{2}{c}{0.0}	        	  \\
$\chi^{2}$                	  	                & 0.008 & 0.003	  \\
 \hline
\end{tabular}
    \label{tab:lcparameters}
     \end{center}
     \centering
     \footnotesize{$^{*}$ fractional radii calculated from the geometric mean $r_{\rm mean}=(r_{\rm pole} \times r_{\rm side} \times r_{\rm back})^{1/3}$}
\end{table}

%Figure 3

%Figure 4

%Figure 5

\section{Fundamental Parameters of IY Aur}
The analysis of the light curve data enabled the estimation of the fundamental parameters of the IY Aur binary system's components. The effective temperature and mass of the primary component were adopted as $T_{\rm 1,eff}=17290 \pm 1600$ K and $M_{1}=6.51 \pm 0.81\,M_{\odot}$ based on the findings of \citet{Khalatyan24} using {\it Gaia} DR3 data. These values served as initial reference points for further calculations. The mass of the secondary component was subsequently derived using the mass ratio obtained from a detailed photometric analysis of the system. To determine the radii of both components, the semi-major axis of the system was calculated using Kepler’s third law, and the fractional radii values were directly taken from the results presented in Table~\ref{tab:lcparameters}. Following this, the luminosities and bolometric magnitudes of the components were computed using the standard solar parameters of $T_{{\rm eff, \odot}} = 5777$ K, $M_{{\rm bol},\odot} = 4.74$ mag, together with the bolometric corrections provided by \citet{Eker20}. The surface gravity values for the primary and secondary components were determined as log\,$g_{1} = 4.01 \pm 0.02$ and log\,$g_{2}=3.49 \pm 0.03$ in cgs units, respectively. The $V$-band interstellar extinction for IY Aur, $A_{\rm V,d}=0.784$ mag, was determined using the same method explained in detail by \citet{Alan25}. The distance to IY Aur was calculated as $d=1690\pm237$ pc, incorporating an interstellar extinction value ($A_{\rm V,d}$), the apparent magnitude of the system, the component light ratios listed in Table~\ref{tab:lcparameters}, and the values $BC_{1}=-1.43$ mag and $BC_{2}=-0.14$ mag determined according to \citet{Eker20}. A detailed summary of the estimated fundamental parameters is presented in Table~\ref{tab:estparameters}.

%\begin{figure}[!ht]
%\begin{center}
%\includegraphics*[scale=0.55,angle=000]{fig5.png}
%\caption{The positions of the primary (blue dot) and secondary (red dot) components of IY Aur in the log\,$L$ - log\,$T$ plane. Evolutionary tracks according to metallicity value of [Fe/H] = 0.5 are depicted by blue and red lines for the primary and secondary component stars, respectively. Theoretical evolutionary curves were computed by MESA code.} \label{fig:Roche_geometry}
%\end{center}
%\end{figure}

% TABLE 4
\begin{table*}
\begin{center}
\centering
\caption{The estimated fundamental stellar parameters for IY Aur.}
\begin{tabular}{lrr}
\hline
  Parameter 			              & Value			\\	
\hline
$M$$_{1}$ ($M_\odot$)	          	&6.51 $\pm$ 0.81   \\	
$M$$_{2}$ ($M_\odot$)	         	&5.39 $\pm$ 0.87   \\
$R$$_{1}$ ($R_\odot$)	          	&4.15 $\pm$ 0.20   \\
$R$$_{2}$ ($R_\odot$)		  		&6.88 $\pm$ 0.33   \\
$a$ ($R$$_{\odot}$)        	   	    &19.04 $\pm$ 0.89   \\
$\log L$$_{1}$ ($L_\odot$)		  	&3.14 $\pm$ 0.20   \\
$\log L$$_{2}$ ($L_\odot$)		  	&2.50 $\pm$ 0.22   \\
$\log g$\,$_{1}$ (cgs)              &4.01 $\pm$ 0.02   \\
$\log g$\,$_{2}$ (cgs)              &3.49 $\pm$ 0.03   \\
$M_{\rm Bol, 1}$ (mag)              &-3.12 $\pm$ 0.51   \\
$M_{\rm Bol, 2}$ (mag)	 	        &-1.52 $\pm$ 0.55   \\
$M_{\rm V, 1}$ (mag)	         	&-1.69 $\pm$ 0.54   \\
$M_{\rm V, 2}$ (mag)	          	&-1.38 $\pm$ 0.58   \\
$A_{\rm V,d}$ (mag)	         	    &0.784 $\pm$ 0.020 \\
$d$ (pc)                            &1690 $\pm$ 237  	   \\
 \hline
\end{tabular}
\label{tab:estparameters}
     \end{center}
\end{table*}

\section{Discussion and Conclusion}

Semi-detached binary stars represent a critical phase in the evolution of close binary systems. In this phase, one of the components fills its Roche lobe and transfers mass to its companion, while the other component remains within its Roche lobe. These systems provide important implications for understanding complex physical processes such as mass transfer, angular momentum loss, magnetic activity, and the combined effects of these processes on the structural and evolutionary properties of the components. Moreover, as a result of the mass transfer, the initially less massive secondary component gradually gains a significant amount of material and eventually becomes the more massive star in the system, taking on the role of the primary component. This phenomenon reveals a key process in binary star evolution known as "mass reversal". Studying semi-detached systems over a wide range of ages is therefore essential for understanding how binary interaction shapes stellar evolutionary tracks, particularly in systems with different initial masses and orbital configurations.

In this research, for the first time, a simultaneous analysis of high-precision {\it TESS} photometric data combined with ground-based multicolor observations has been performed for the IY Aur binary system. The system exhibits the characteristics of an evolved semi-detached configuration, in which the originally less massive secondary component has gained mass from the primary, becoming the more massive component in the system, while both components are still undergoing main-sequence evolution. The analysis revealed that the secondary component fills its Roche lobe, while the primary component remains within its critical equipotential surface, with a fill-out factor of $f=65\%$, confirming the semi-detached nature of the system. The solutions indicate a significant temperature difference between the components. Considering that the system is relatively far from the contact phase, this may suggest ongoing or recent mass transfer activity. The hotspot revealed through the light curve analysis may also be interpreted as an indirect indicator of ongoing mass transfer within the system. In binary systems composed of early-type components, the presence of such hot spots is generally associated with thermal enhancements occurring in localized regions where the accretion stream impacts the surface of the mass-accreting component with high velocity during the Roche lobe overflow phase. This interaction leads to transient temperature anomalies on the stellar surface, producing detectable signatures in both photometric and spectroscopic observations. The dynamical interaction between the mass stream and the stellar surface would potentially reveal more complex features. However, the current dataset does not exhibit clear evidence of such interactions beyond what can be reasonably modeled with a hot spot, suggesting either a low mass transfer rate or that the geometry of the stream does not result in prominent photometric effects. A more detailed understanding of the dynamic properties of the mass flow, its physical structure, and the interaction region between the components can only be achieved through high-resolution spectroscopic observations in which the spectral components of the system are disentangled.

The fundamental parameters of each component of the IY Aur system have been determined, providing valuable data for stellar astrophysics. The masses of the components were derived as $M_{1}= 6.51\pm 0.81\,M_\odot$ for the primary component and $M_{2}= 5.39\pm0.87\,M_\odot$ for the secondary component. The total mass of the system, which is about 12 $M_\odot$, is an uncommon feature among close binary systems that have undergone a mass transfer process. The radii were found to be $R_{1}=4.15\pm 0.20\,R_\odot$ and $R_{2}= 6.88\pm 0.33\,R_\odot$, respectively. The derived logarithmic luminosities and surface gravities are $\log L_{1}$=3.14 $\pm$ 0.20 $L_\odot$ and $\log g_{1}$=4.01 $\pm$ 0.02 cgs for the primary star, and $\log L_{2}$=2.50 $\pm$ 0.22 $L_\odot$ and $\log g_{2}$=3.49 $\pm$ 0.03 cgs for the secondary star.

Following the determination of the fundamental parameters, the distance of IY Aur to our Solar System was calculated as $d=1690\pm 237$ pc, which is slightly smaller than the {\it Gaia} DR3 distance estimate of 1919 pc. The discrepancy may arise from the complex structure of the system and the presence of mass transfer activity, which can affect astrometric measurements in binary systems.

In conclusion, the IY Aur system, which contains relatively massive components, currently in a semi-detached evolutionary phase, offers a valuable laboratory for investigating the effects of mass transfer and binary interaction on stellar structure and evolution. For a more comprehensive understanding of the system's nature and its evolutionary status, future high-resolution spectroscopic studies are essential. Such data will provide detailed information on the chemical composition, rotational velocities, and orbital dynamics of the components, further contributing to the broader field of close binary star astrophysics.

\section*{Acknowledgements}

This research was funded by the Scientific Research Projects Coordination Unit of Istanbul University under project number 37903. We extend our gratitude to the T\"{U}B\.{I}TAK National Observatory (TUG) for partially supporting the use of the T60 telescope through project number 18BT60-1324. We also appreciate the invaluable contributions of the observers and technical staff at TUG, both before and during the observations.

This study made use of National Aeronautics and Space Administration's (NASA) Astrophysics Data System, as well as the SIMBAD Astronomical Database, operated by the Centre de Données astronomiques de Strasbourg (CDS) in France. Additionally, data from the NASA/IPAC Infrared Science Archive, managed by the Jet Propulsion Laboratory at the California Institute of Technology under a contract with NASA, were utilized. Furthermore, this research incorporated data from the European Space Agency (ESA) mission $Gaia$\footnote{https://www.cosmos.esa.int/gaia}, processed by the $Gaia$ Data Processing and Analysis Consortium (DPAC)\footnote{https://www.cosmos.esa.int/web/gaia/dpac/consortium}, with funding provided primarily by national institutions participating in the Gaia Multilateral Agreement. The {\it TESS} data used in this study were retrieved from the Mikulski Archive for Space Telescopes (MAST), with the {\it TESS} mission being funded by NASA’s Explorer Program.

%%%%%%%%%%%%%%%%%%%% REFERENCES %%%%%%%%%%%%%%%%%%
% The best way to enter references is to use BibTeX:

\bibliographystyle{mnras}
\bibliography{Alan_et_al} % if your bibtex file is called example.bib

%\bibliographystyle{ieeer}

% Don't change these lines
\bsp	% typesetting comment
\label{lastpage}
\end{document}